\documentclass[runningheads,a4paper]{llncs}

\usepackage{mathptmx}       % selects Times Roman as basic font
\usepackage{helvet}         % selects Helvetica as sans-serif font
\usepackage{courier}        % selects Courier as typewriter font
\usepackage{type1cm}        % activate if the above 3 fonts are
\usepackage{makeidx}         % allows index generation
\usepackage{graphicx}        % standard LaTeX graphics tool
\usepackage[bottom]{footmisc}% places footnotes at page bottom
\makeindex             % used for the subject index
\usepackage{lscape} 
\usepackage{subfigure} 
\newcommand\apjl{ApJL }     % Astrophysical Journal, Letters 
\newcommand\apjs{ApJS }%    % Astrophysical Journal, Supplement 
\newcommand\aap{A\&A }%     % Astronomy and Astrophysics 
\newcommand{\Teff}{$T_{\rm eff}$}
\newcommand{\logg}{log\,\textit{g}}
\newcommand{\vsini}{\textit{v}\,sin\,\textit{i}}

\newcommand{\Mdot}{\textit{\.M}}
\DeclareRobustCommand\ion[2]{%
  \mbox{#1\kern0.2em%
  \smaller\rmfamily%
  \edef\@tempa{\@car#2\@nil}%
  \ifcat1\@tempa%
  \@Roman{#2}%
  \else%
  \uppercase{#2}%
  \fi}}
  \newcounter{IonStage}

\def\ion#1#2{\ensuremath{\mathrm{#1}\;}{\protect\small\rm{#2}}}
\DeclareRobustCommand\nodata{$\cdots$}
\begin{document}

\title{Creating and using large grids of pre-calculated model atmospheres for rapid 
analysis of stellar spectra}
\titlerunning{Large grids of model atmospheres}
\author{Janos Zsarg\'o\inst{1}, Celia Rosa Fierro\inst{2}, Jaime Klapp\inst{2}, Anabel
  Arrieta\inst{3}, Lorena Arias\inst{3}, Jurij Mendoza Valencia\inst{1}, Leonardo Di G. Sigalotti\inst{4}} 
  \institute{Departamento de
  F\'{i}sica, Escuela Superior de F\'{i}sica y Matem\'aticas,
  Instituto Polit\'ecnico Nacional, Ciudad de M\'{e}xico, M\'{e}xico,\\ 
  \and
  Instituto Nacional de Investigaciones Nucleares (ININ), Estado de M\'exico, M\'{e}xico,\\
  \and
  Universidad Iberoamericana, Ciudad de M\'{e}xico, M\'{e}xico,
 \and
  \'Area de F\'isica de Procesos Irreversibles, Departamento de Ciencias B\'asicas,
  Universidad Aut\'onoma Metropolitana-Azcapotzalco (UAM-A), Ciudad de M\'{e}xico, 
  M\'{e}xico.\\
 \email{jzsargo@esfm.ipn.mx},
 \email{celia.fierro.estrellas@gmail.com},
 \email{jaime.klapp@inin.gob.mx},
 \email{anabel.arrieta@uia.mx},
 \email{lorena.arias@uia.mx},
 \email{jurijmev@gmail.com}
 \email{leonardo.sigalotti@gmail.com}
}

\authorrunning{Zsarg\'o et al.}
 
\thispagestyle{empty} \maketitle \thispagestyle{empty}
\setcounter{page}{123}

\abstract{We present a database of 45,000 atmospheric models (which will become
  80,000 models by the end of the project) with stellar masses between 9 and 
  120\,M$_{\odot}$, covering the region of the OB main sequence and W-R stars in 
  the H--R diagram. The models were calculated using the ABACUS I supercomputer 
  and the stellar atmosphere code CMFGEN. The parameter space has 6 dimensions: the
  effective temperature \Teff, the luminosity $L$, the metallicity $Z$, and three 
  stellar wind parameters, namely the exponent $\beta$, the terminal velocity 
  $V_{\infty}$, and the volume filling factor $F_{cl}$. For each model, we also 
  calculate synthetic spectra in the UV (900-2000 \AA), optical (3500-7000 \AA),
  and near IR (10000-30000 \AA) regions. To facilitate comparison with observations, 
  the synthetic spectra were rotationally broaden using ROTIN3, by covering \vsini\ 
  velocities between 10 and 350 km/s with steps of 10\,km/s, resulting 
  in a library of 1 575 000 synthetic spectra. In order to demonstrate the benefits
  of employing the databases of pre-calculated models, we also present the results 
  of the re-analysis of $\epsilon$~Ori by using our grid.}


\section{Introduction}
\label{sec:intro}

Thanks to the fertile combination of the large amount of public data and the availability 
of sophisticated stellar atmosphere codes such as CMFGEN \cite{hil98}, TLUSTY \cite{hub95}, 
FASTWIND \cite{san97,pul05}, and the Potsdam Wolf-Rayet code (PoWR) \cite{gra02,ham04} 
self-consistent analysis of spectral regions from the UV to the IR is now possible. As a 
result of this we have made significant advances in the understanding of the physical 
conditions in the atmospheres and winds of massive stars. 

For example, early far-UV observations showed that there were inconsistencies between the 
optical effective temperature scale and that implied by the observed wind ionization 
\cite{ful00}. Studies by Martins et al. \cite{mar02}, and others, have shown that the 
neglect of line blanketing in the models leads to a systematic overestimate of the 
effective temperature when derived from optical H and He lines. On the other hand,
Crowther et al. \cite{cro02}, Bouret at al. \cite{bou03}, and Hillier et al. \cite{hil03} 
simultaneously analyzed the $FUSE$, the $HST$, and the optical spectra of O stars and were 
able to derive consistent effective temperatures using a wide variety of diagnostics. 

Another crucial result was the recognition of the important effect of wind inhomogeneities 
(clumping) on the spectral analyses of O stars. For instance, Crowther et al. \cite{cro02} 
and Hillier et al. \cite{hil03} could not reproduce the observed \ion{P}{V} 
$\lambda\lambda$1118-1128 profiles when using mass-loss rates derived from the analysis of 
H$\alpha$ lines. The only ways the \ion{P}{V} and the H$\alpha$ profile discrepancies could 
be resolved were either by assuming substantial clumping or using unrealistically low 
phosphorus abundances. As a consequence of clumping, the mass-loss rates have been lowered 
by significant factors (i.e., from 3 to 10). However, the possibility of optically thick 
clumping was raised recently which would change this conclusion (see, e.g., Ref. \cite{sun14} 
and references therein).

Unfortunately, performing such investigations by using any of the above mentioned stellar 
atmosphere codes is not an easy task! To run these codes and perform a reliable analysis 
requires a lot of experience; something that many investigators do not have the time to gain. 
Therefore, it is useful to develop databases of pre-calculated models. Such databases will 
free up valuable time for astronomers, who could study stellar atmospheres with reasonable 
accuracy but without the need of running time consuming simulations. Furthermore, these 
databases will also accelerate the studies of large numbers of observed spectra that are in 
line for analysis.  

The basic parameters of such databases of pre-calculated models are: the surface temperature 
(\Teff), the stellar mass ($M$), and the surface chemical composition. An adequate
analysis of massive stars also has to take into account the parameters associated with the 
stellar wind, such as the terminal velocity ($V_\infty$), the mass-loss rate ($\Mdot$), and 
the clumping. If one takes into account the variations of all necessary parameters the number 
of pre-calculated models that are needed increases exponentially. Therefore, production of 
such databases is only possible by using supercomputing facilities.

There are already a few databases of synthetic stellar spectra available, but only with a few 
tens or hundreds of stellar models (see, for example, the atlas of CMFGEN models for OB
massive stars by Fierro et al. \cite{fie15}, the grid of W-R stars by Hamann and Gr\"afener
\cite{ham04}, and the POLLUX database by Palacios et al. \cite{pal10}). On the other hand, 
we are generating a database with tens of thousands of models \cite{zsa17}, which will be 
publicly available in a year or so. Obviously, it will be impossible to manually compare an 
observed spectrum with such an amount of model calculations. Therefore, it is imperative to 
develop tools that allow the automation of this process but without compromising the quality 
of the fit. In particular, in Ref. \cite{fie18} we have presented FIT\textit{spec}, a 
program that searches our database for a model that better fits the observed spectrum in the 
optical region. It uses the Balmer lines to measure the surface gravity ($\log (g)$) and the 
equivalent width ratios of \ion{He}{II} and \ion{He}{I} lines to estimate \Teff. 

In this article we describe the state of our grid of pre-calculated models and the results 
of a test analysis to verify the usefulness of the grid. In \S~\ref{sec:cmfgen} and 
\S~\ref{sec:cmf_flux}, we briefly describe the stellar atmosphere code (CMFGEN) which we 
use to produce our models. In \S~\ref{sec:models}, we describe our model grid and in 
\S~\ref{sec:simtest} we describe a simple test analysis to demonstrate the benefits of 
using our grid. Finally, in \S~\ref{sec:concl} we summarize the relevant conclusions.

\section{CMFGEN}
\label{sec:cmfgen}

CMFGEN is a sophisticated and widely-used non-LTE stellar atmosphere code \cite{hil03,hil98}. 
It models the full spectrum and has been used successfully to model O \& B stars, W-R stars, 
luminous blue variables, and even supernovae. The code determines the temperature,
ionization structure, and level populations for all elements in the stellar atmosphere and 
wind. It solves the spherical radiative transfer equation in the co-moving frame in conjunction
with the statistical equilibrium equations and radiative equilibrium equation. The hydrostatic 
structure can be computed below the sonic point, thereby allowing the simultaneous treatment 
of spectral lines formed in the atmosphere, the stellar wind, and the transition region 
between the two. Such features make it particularly well suited for the study of massive OB 
stars with winds. However, there is a price for such sophistication, a CMFGEN simulation 
takes anywhere between 24 and 36 hours of microprocessor time to be completed.

For atomic models, CMFGEN utilizes the concept of ``super levels'' by which levels of similar 
energies are grouped together and treated as a single level in the statistical equilibrium 
equations (see, Ref. \cite{hil98} and references therein for more details). The stellar models 
in this project include 28 explicit ions of the different elements as function of their \Teff. 
Table~\ref{tab:superlevels} summarizes the levels and super levels included in the models. 
The atomic data references are given by Herald and Bianchi \cite{her04}.
%
%\begin{landscape} 
\begin{table}
\begin{center}
\caption{Super levels/levels for the different ionization stages included in the models.
\label{tab:superlevels}}
\begin{tabular}{lllllllll}
%\tableline\tableline
\multicolumn{1}{c}{Element} &\multicolumn{1}{c}{I} &
\multicolumn{1}{c}{II} &\multicolumn{1}{c}{III}&\multicolumn{1}{c}{IV}& 
\multicolumn{1}{c}{V} &\multicolumn{1}{c}{VI}&\multicolumn{1}{c}{VII}&
\multicolumn{1}{c}{VIII}\\
\hline
H & 20/30 & 1/1   &\nodata&\nodata&\nodata&\nodata&\nodata&\nodata\\
He& 45/69 & 22/30 & 1/1   &\nodata&\nodata&\nodata&\nodata&\nodata\\
C &\nodata& 40/92 & 51/84 & 59/64 &  1/1  &\nodata&\nodata&\nodata\\
N &\nodata& 45/85 & 41/82 & 44/76 & 41/49 &  1/1  &\nodata&\nodata\\
O &\nodata& 54/123& 88/170 & 38/78 & 32/56& 25/31 &  1/1  &\nodata\\
Si&\nodata&\nodata& 33/33 & 22/33 & 1/1   &\nodata&\nodata&\nodata\\
P &\nodata&\nodata&\nodata& 30/90 & 16/62 &  1/1  &\nodata&\nodata\\
S &\nodata&\nodata& 24/44 & 51/142& 31/98 & 28/58 &  1/1  &\nodata\\
Fe&\nodata&\nodata&104/1433&74/540& 50/220& 44/433 & 29/153&  1/1 \\ 
\hline
\end{tabular}
\end{center}
\end{table}
%\end{landscape}
%%\clear
 
To model the stellar wind, CMFGEN requires values for the mass loss rate (\Mdot), terminal 
velocity ($V_{\infty}$), $\beta$ parameter, and the \textit{volume filling factor} of the 
wind ($F_{cl}$). The profile of wind speed is modeled by a beta-type law \cite{CAK}

\begin{equation}
\label{e:beta_law}
\mathbf{v} (\mathbf {r})= v_{\infty}\left(1- \frac{r}{R_{*}}\right)^{\beta},
\end{equation}

\begin{figure*}
\centering
\includegraphics[width=0.49\linewidth]{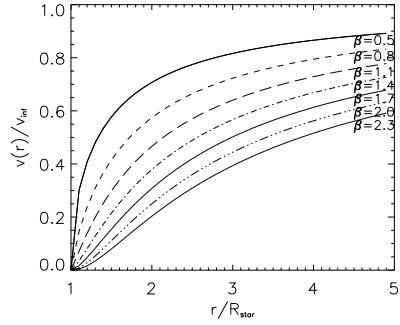}
\caption{Examples of beta-type velocity laws.}
\label{f:beta_law}
\end{figure*}

The $\beta$ parameter controls how the stellar wind is accelerated to reach the terminal 
velocity (see Fig.~\ref{f:beta_law}), while the volume filling factor $F_{cl}$ is used to 
introduce the effects of optically thin clumping in the wind (see Ref. \cite{sun14} and 
references therein). 
%
% FIGURE  beta law
%

\subsection{Synthetic spectra}
\label{sec:cmf_flux}

The auxiliary program CMF\_FLUX of the CMFGEN package computes the synthetic observed
spectrum in the observer's frame which is one of the most important output of our models 
\cite{hil13}. To simulate the effects of rotation on the spectral lines, the synthetic 
spectra are also rotationally broadened using the program ROTIN3, which is part of the 
TLUSTY package \cite{hub95}.

For each model in the grid, we calculate the normalized spectra in the UV (900-3500
\AA\hspace{0.1cm}), optical (3500-7000 \AA\hspace{0.1cm}), and IR (7000-40 000 
\AA\hspace{0.1cm}) range; then, we apply rotation by sampling the range between 10 and 
350 km/s$^{-1}$ with steps of 10\ km/s$^{-1}$. This process results in a library with 
a total of 1 575 000 synthetic spectra.

\section{The model grid}
\label{sec:models}

The main parameters of a model atmosphere are the luminosity (\textit{L}) and the 
effective temperature (\Teff), whose values allow to place the star in the H-R diagram. 
In order to constrain appropriately the input parameters, we use the evolutionary tracks 
of Ekstr\"om et al. \cite{eks12} calculated with solar metallicity (Z=0.014) at the zero 
age main sequence (ZAMS). For any track, each point corresponds to a star with specific 
values of \Teff, luminosity ($L$), and stellar mass ($M$). We calculated several models 
along each track with the approximate steps of 2\,500\,K in \Teff, while the stellar 
radius and the surface gravity $\log (g)$ were calculated to get the luminosity $L$ and 
the stellar mass $M$ corresponding to the track.

The elements included in our models are H, He, C, N, O, Si, P, S, and Fe. The values of 
H, He, C, N, and O were taken from the tables of Ekstr\"om et al. \cite{eks12}. For
consistency, we assumed solar metallicity as reported by Asplund et al. \cite{asp09} for
Si, P, S, and Fe in all models. 

\begin{figure*}
%\centering
\includegraphics[width=0.8\textwidth]{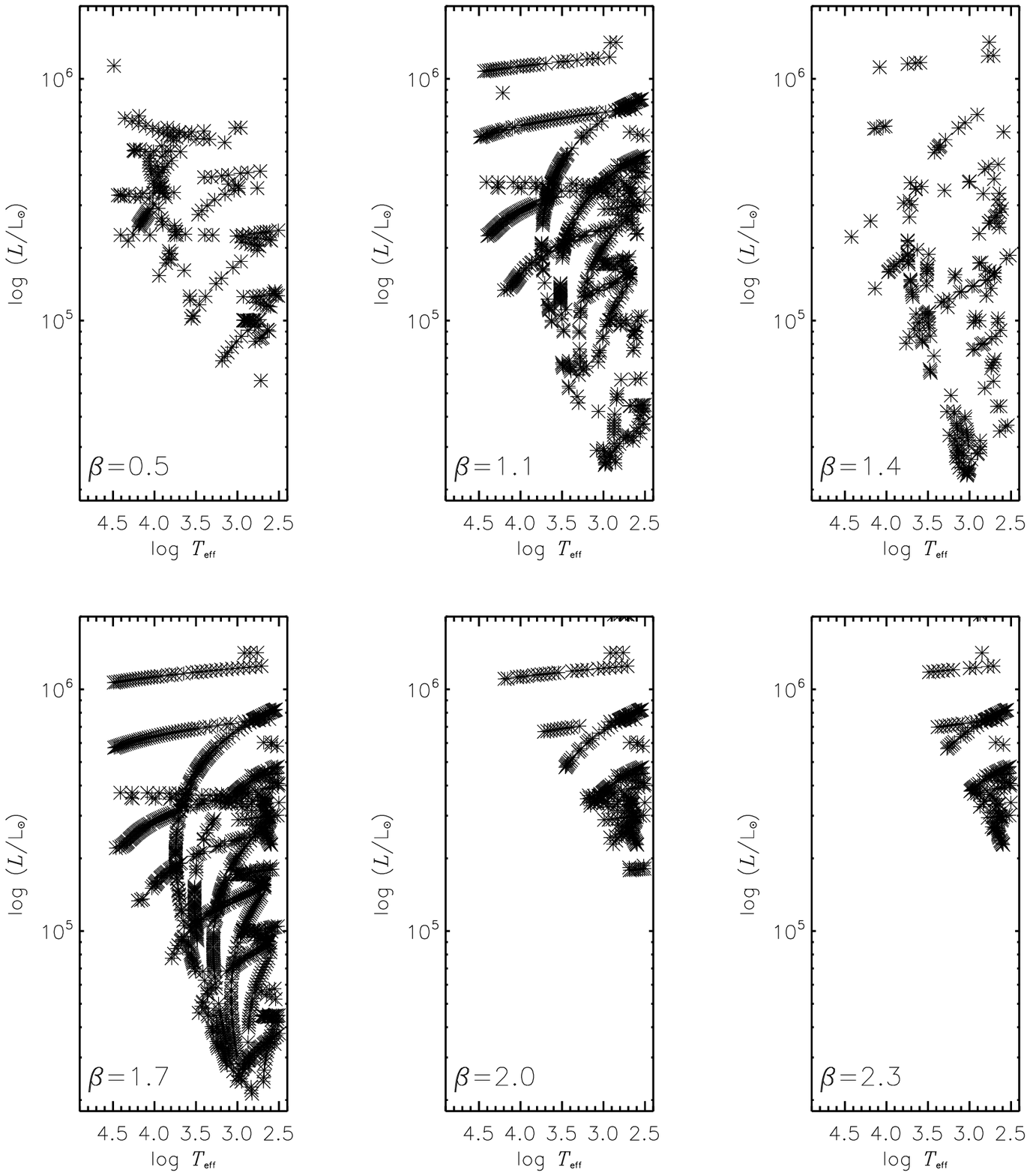}
\includegraphics[width=0.45\textwidth,bb=0 0 533 462]{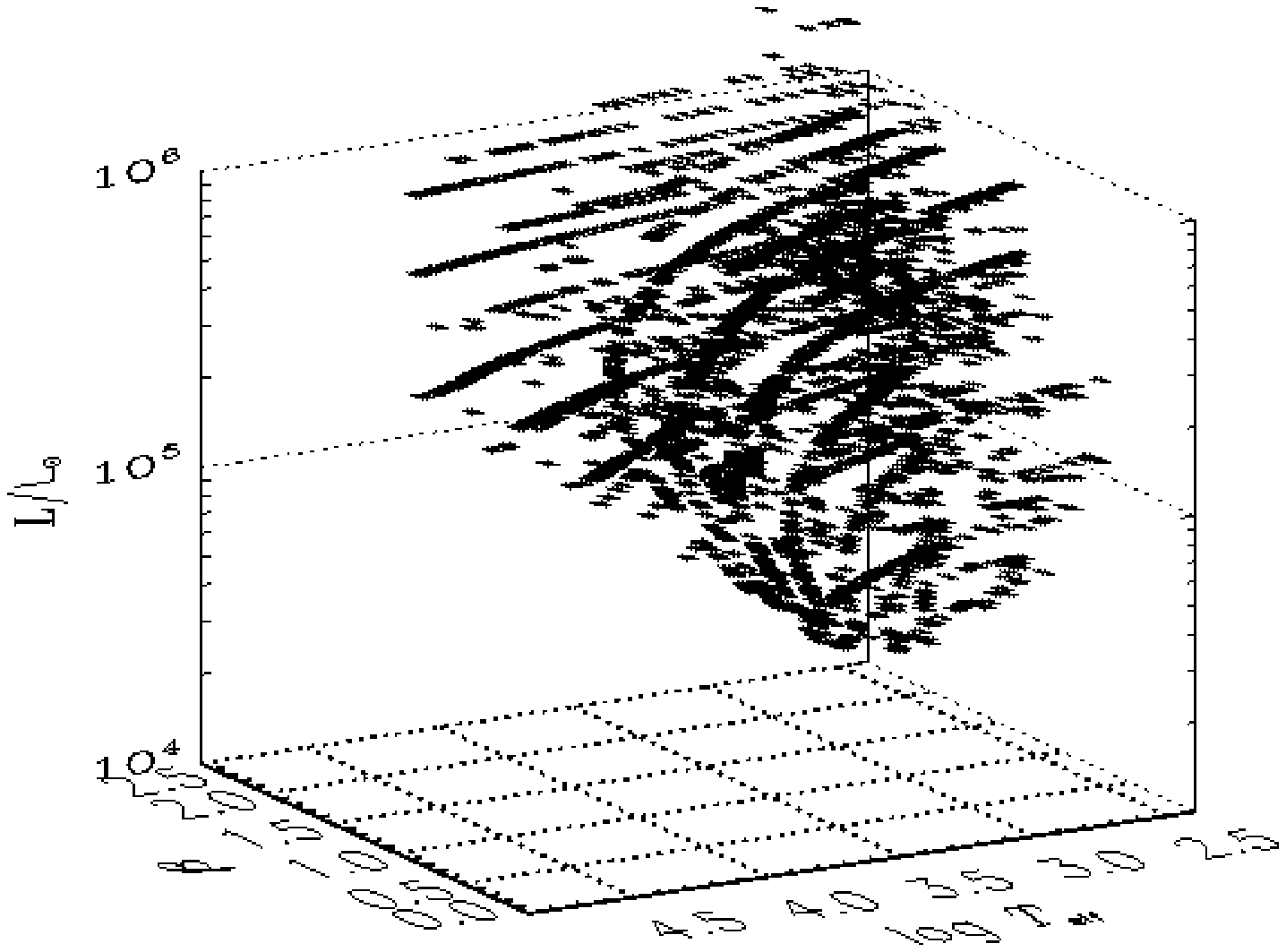}
\includegraphics[width=0.45\textwidth,bb=0 0 633 462]{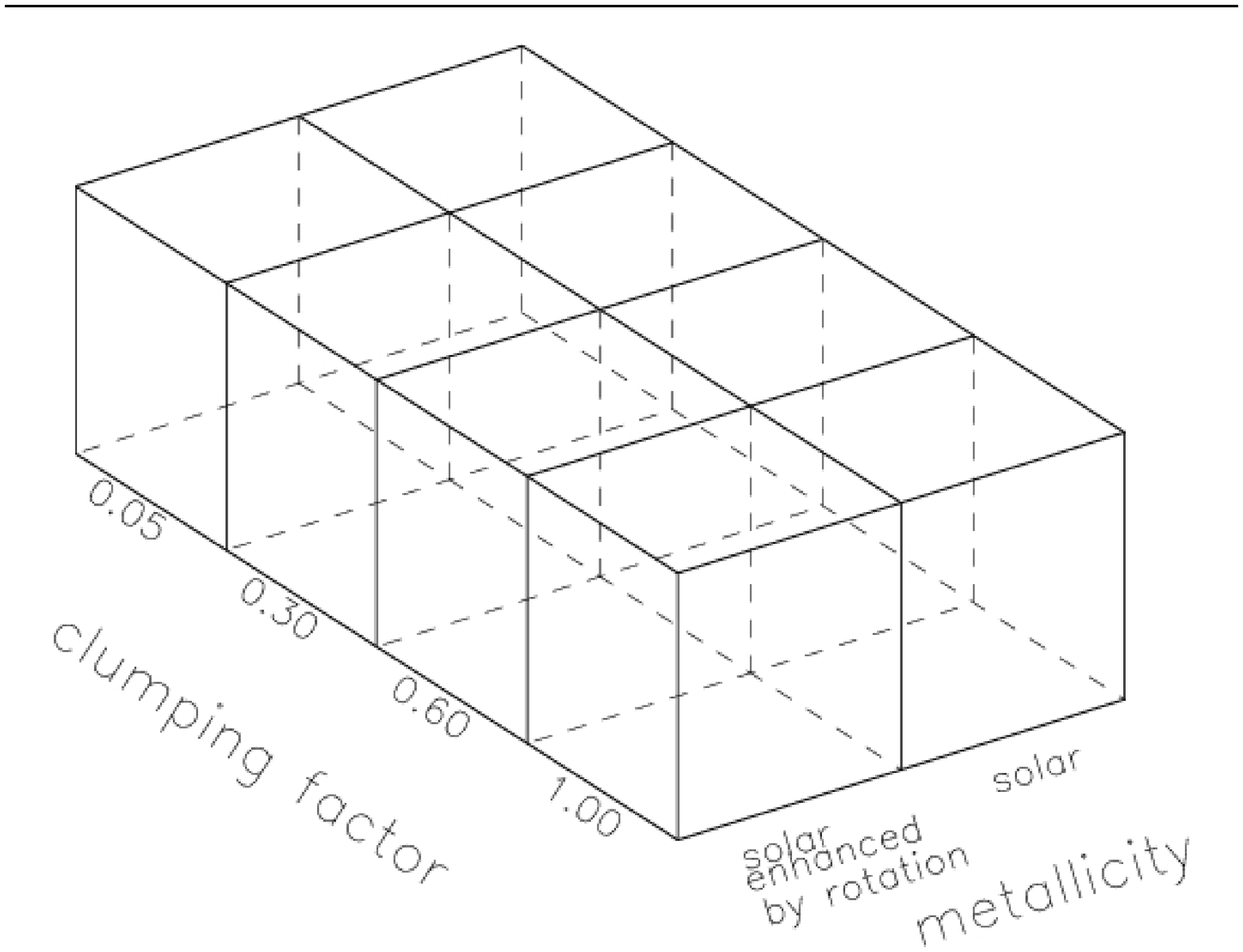}
\caption{Organization of the grid as a 5-dimensional hypercube.
\textit{Top:}  \Teff-Luminosity planes with different values of $\beta$ parameter.
\textit{Bottom Left:} Data cube with the models contained in the six planes.
\textit{Bottom Right:} Plane formed by cubes similar to that shown on the left, the
 dimensions of these are different values of the \textit{volume filling factor} 
with two different metallicities.}
\label{f:grid}
\end{figure*}

The grid is organized as a hypercube data in dimensions which correspond to \Teff, $L$, 
$V_{\infty}$, $\beta$, $F_{cl}$, and the metallicity. The plane generated by 
\Teff\hspace{0.1cm} and $L$ is the H-R diagram (see upper part of Figure~\ref{f:grid}); 
the values of these variables are restricted by evolutionary tracks. For $V_{\infty}$ we 
use two values, a low ($V_{\infty}=1.3V_{esc}$) and a high ($V_{\infty}=2.1V_{esc}$) 
velocity model, where the escape velocity ($V_{esc}$) has the usual meaning. The fourth 
dimension is the $\beta$ parameter of the stellar wind for which we use the values of 
$\beta=0.5$, 0.8, 1.1, 1.4, 1.7, 2.0, and 2.3 (see bottom left of Figure~\ref{f:grid}). 
Models with different values of \Teff, $L$, and $F_{cl}$ populate a data cube. Each value 
of $F_{cl}=0.05$, 0.30, 0.60, and 1.0 generates a similar cube, all of which are aligned 
one after another in a fifth dimension. Finally, we have two values of metallicity: solar 
and solar enhanced by rotation. This 6-dimensional arrangement generates a plane populated 
with data cubes (see bottom right of Figure~\ref{f:grid}).

This arrangement only populates regions of the H-R diagram where nature forms stars, 
and does not produce non-physical models. If needed, we can interpolate between models 
to achieve better fits to the observed spectra.

\section{A simple test to demonstrate the usefulness of our grid}
\label{sec:simtest}

\begin{figure*}
\centering
\includegraphics[width=0.9\textwidth]{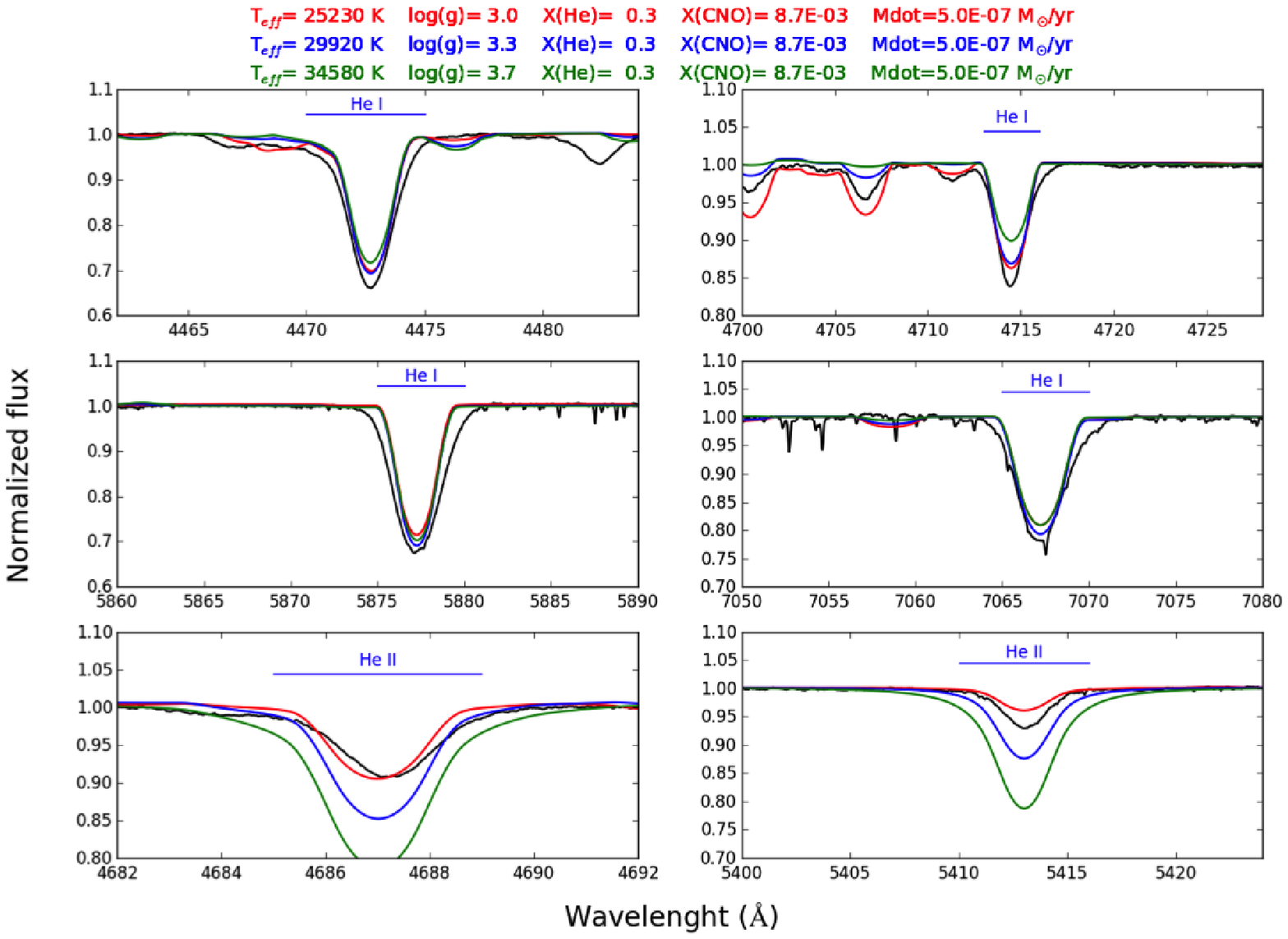}
\caption{Comparison of selected models (coloured lines) with the observed \ion{He}{I} and 
\ion{He}{II} lines for $\epsilon$~Ori (black curves) to estimate the $T_{eff}$ of the star. 
The parameters of the models are colour-coded above the figure.}
\label{f:Teff}
\end{figure*}

We demonstrate the benefits of having a mega-grid by a re-analysis of $\epsilon$~Ori. This 
O9/B0 supergiant was studied by Puebla et al. \cite{pue16} by using CMFGEN in the traditional 
way (i.e., by producing every model that was needed). They reported $T_{eff}=$27,000~K, \logg\ 
= 3, a mass-loss rate \Mdot $\sim 10^{-7} M_{\odot}$ yr$^{-1}$, and a highly clumped and 
slowly accelerating wind ($F_{cl}=0.01$, $\beta >2.0$) for this star.
Figure~\ref{f:Teff} shows a comparison of selected models from the grid with the optical 
\ion{He}{I} and \ion{He}{II} lines observed for $\epsilon$~Ori. We tried to select models 
which only differ in the effective temperature and have a low mass-loss rate to avoid 
complications. Obviously, our grid is still not fine enough to be able to do that.

The He lines are normally used to estimate the effective temperature of O stars. Although, 
the \ion{He}{II} lines are very weak for $\epsilon$~Ori since this star is on the borderline 
between type O and B, the comparison still shows that $T_{\rm eff}$ has to be around 25,000~K. 

\begin{figure*}
\centering
\includegraphics[width=0.9\textwidth]{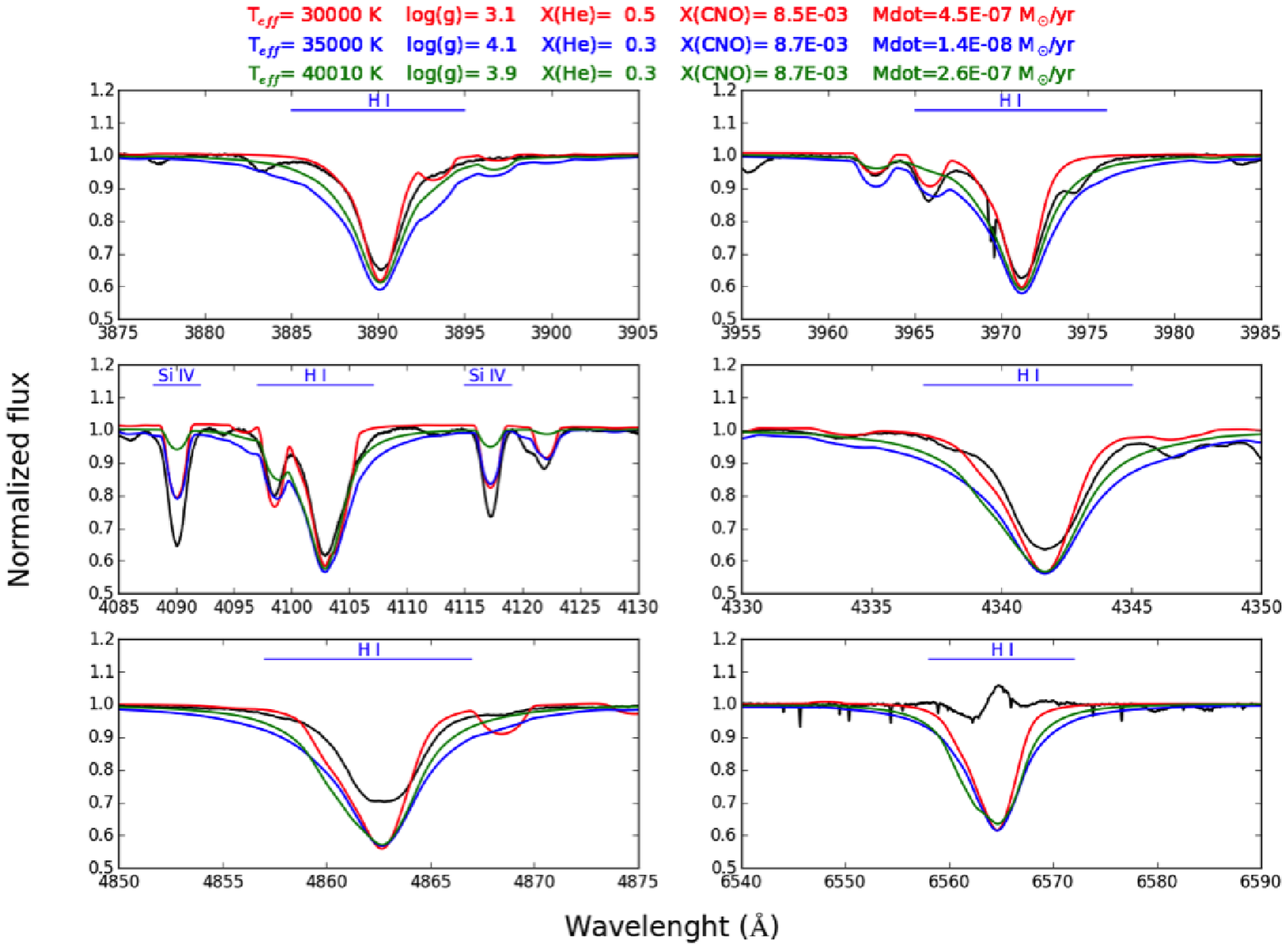}
\caption{Comparison of selected models (coloured lines) with the observed \ion{H}{I} Balmer 
series for $\epsilon$~Ori (black curves) to estimate the surface gravity (\logg) of the star. 
The parameters of the models are colour-coded above the figure.}
\label{f:logg}
\end{figure*}

Moreover, Figure~\ref{f:logg} shows models from the grid which vary in \logg\ in comparison 
with the observed H Balmer series. Unfortunately, the H Balmer lines are also affected by 
mass-loss, namely the absorption is filled in by the emission in the base of the wind. For 
example, Fig.~\ref{f:logg} shows that H$\alpha$ and H$\beta$ are basically useless as 
\logg\ indicator even at relatively low \Mdot. Neverheless, the higher order members of the 
Balmer series are not affected by mass-loss and they support the published value of 
\logg\ $\sim$3.

\begin{figure*}
\centering
\includegraphics[width=0.9\textwidth]{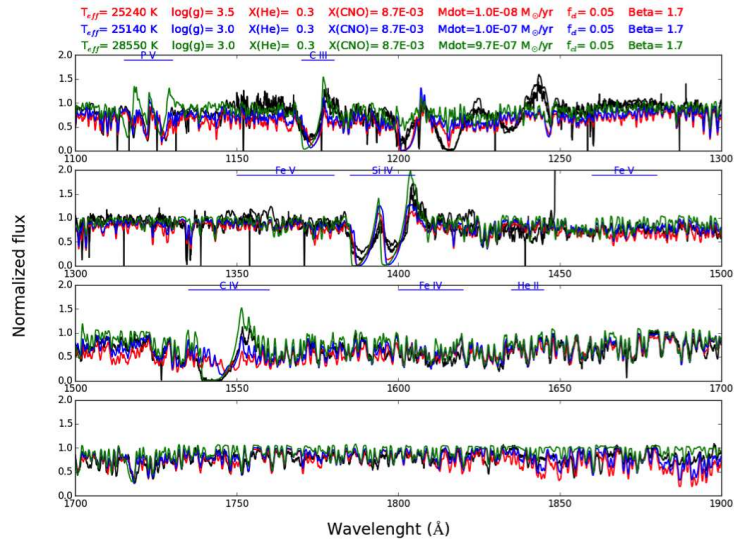}
\caption{Comparison of selected models (coloured lines) with the observed UV spectra ($IUE$) 
for $\epsilon$~Ori (black curves) to estimate the mass-loss rate (\Mdot) of the star. The 
parameters of the models are colour-coded above the figure.}
\label{f:UV}
\end{figure*}

The most useful spectral region to estimate the mass-loss rate (\Mdot) and terminal velocity 
($V_{\infty}$) is the ultraviolet one. Here, we encounter strong resonance lines of the 
dominant ionization states of various elements in the winds of massive stars. These lines 
normally show P-Cygnii profiles which are particularly useful to measure \Mdot\ and 
$V_{\infty}$; see e.g., the \ion{C}{IV} doublet around 1550\AA\ or the \ion{Si}{IV} doublet 
around 1400\AA\ in Fig.~\ref{f:UV}. However, these lines are not useful to estimate $F_{cl}$ 
and, if saturated, they are also useless to measure $\beta$. The comparison of models with 
the observations in Fig.~\ref{f:UV} shows a somewhat contradictory situation, while 
the \ion{Si}{IV} $\lambda\lambda$1400 doublet suggests a low \Mdot. To fit the \ion{C}{IV} 
$\lambda\lambda$1552 profile we would need much higher mass-loss rates. However, 
\Mdot $\sim 10^{-6} M_{\odot}$ yr$^{-1}$ would result in H$\alpha$ emission which is not 
observed. Therefore, we conclude that \Mdot $\sim 10^{-7} M_{\odot}$ yr$^{-1}$ is the best 
estimate we can have.

\begin{figure*}
\centering
\includegraphics[width=0.9\textwidth]{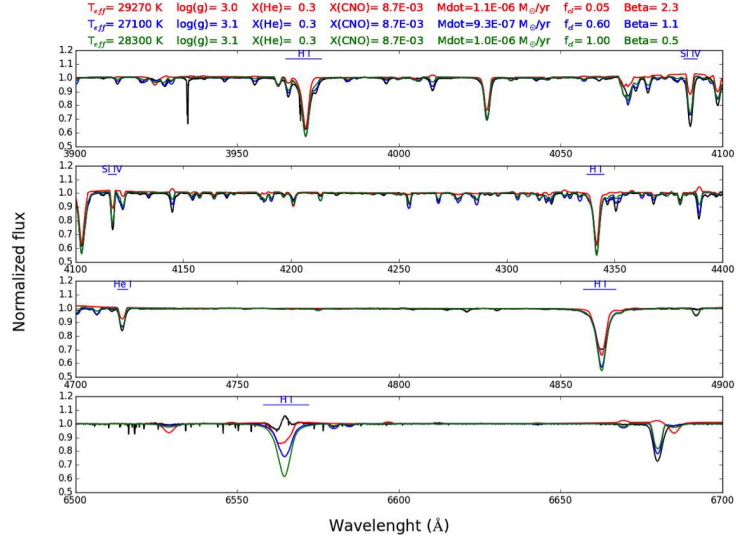}
\caption{Comparison of selected models (coloured lines) with the optical spectrum observed 
for $\epsilon$~Ori (black curves). We can estimate the $F_{cl}$ and $\beta$ (see 
\S~\ref{sec:cmfgen} for an explanation) using H$\alpha$. The parameters of the models are 
colour-coded above the figure.}
\label{f:BetaFcl}
\end{figure*}

Estimations of $F_{cl}$ and $\beta$ are very difficult because we do not have many diagnostics 
and those that we have, like H$\alpha$, are affected by a combination of parameters. 
Nevertheless, the comparison of selected models with the observed H$\alpha$ lines in 
Fig.~\ref{f:BetaFcl} indicates that the values reported in Ref. \cite{pue16} are plausible. 

As expected, the re-analysis of $\epsilon$~Ori supported the values that were published in 
Ref. \cite{pue16}. Obviously, the analysis by using the grid is much cruder, but also much 
faster. We needed only an afternoon to perform the analysis presented here, while several 
months of work was necessary to achieve the results presented by Puebla et al. \cite{pue16}.

\section{Summary}
\label{sec:concl}

We present a mega grid of 45,000 (which will soon becomes more than 80,000) stellar 
atmospheric models calculated by the CMFGEN package. These models cover the region of the 
H-R diagram populated by OB main sequence and W-R stars with masses between 9 and 
120\,$M_{\odot}$. The grid provides UV, visual, and IR spectra for each model. 

We use  \Teff\hspace{0.05cm} and luminosity values that correspond to the evolutionary 
tracks of Ekstr\"om et al. \cite{eks12}. Furthermore, we sample seven values of $\beta$, 
five values of the clumping factor, and two different metallicities and terminal velocities. 
This generates a 6-dimensional hypercube of stellar atmospheric models, which we intend to 
release to the general astronomical community as a free tool for analyzing OB stars.

We have also demonstrated the usefulness of our mega-grid by the re-analysis of 
$\epsilon$~Ori. Our somewhat crude but very rapid analysis supported the stellar an wind 
parameters reported by Puebla et al. \cite{pue16}. The re-analysis has demonstrated the 
benefits of having a large grid of pre-calculated models. This way we can perform rapid
and reliable estimates of the stellar and wind parameters for a star; and if needed, a more 
detailed study can be performed but by starting with good initial values. This significantly 
shortens the time that is needed to complete the spectral analysis.    

\section*{Acknowledgments}
The authors acknowledge the use of the ABACUS-I supercomputer at the Laboratory of Applied Mathematics and High-Performance Computing of the Mathematics Department of CINVESTAV-IPN, where this work was performed.  J. Zsargo acknowledges CONACyT CB-2011-01 No. 168632 grant for support. The research leading to these results has received funding from the European Union's Horizon 2020 Programme under the ENERXICO Project, grant agreement no 828947 and under the Mexican CONACYT-SENER-Hidrocarburos grant agreement B-S-69926. J. K. acknowledges financial support by the Consejo Nacional de Ciencia y Tecnolog\'ia (CONACyT), M\'exico, under grant 283151.

\bibliography{ms}
\bibliographystyle{splncs03}

\end{document}